\documentclass[fleqn,usenatbib]{mnras}

\usepackage{newtxtext,newtxmath}

\usepackage[T1]{fontenc}

\DeclareRobustCommand{\VAN}[3]{#2}
\let\VANthebibliography\thebibliography
\def\thebibliography{\DeclareRobustCommand{\VAN}[3]{##3}\VANthebibliography}

\usepackage{graphicx}	
\usepackage{amsmath}	
\usepackage{amsmath}
\usepackage[multiple]{footmisc}

\title[Higher-Order Clustering]{Information Content of Higher-Order Galaxy Correlation Functions}

\author[Samushia et al.]{
Lado Samushia,$^{1,2\thanks{E-mail: lado@phys.ksu.edu}}$ Zachary Slepian,$^{3,4\thanks{E-mail: zslepian@ufl.edu}}$ \& Francisco Villaescusa-Navarro$^{5,6}$\thanks{E-mail: fvillaescusa@princeton.edu}
\\
$^{1}$ Department of Physics, Kansas State University, 116 Cardwell Hall, 1228 N. 17$^{th}$ St., Manhattan, KS 66506, USA\\
$^{2}$ Abastumani Astrophysical Observatory, Tbilisi, GE-0179, Georgia\\
$^{3}$ Department of Astronomy, University of Florida, 211 Bryant Space Science Center, Gainesville, FL 32611\\
$^{4}$ Physics Division, Lawrence Berkeley National Laboratory, 1 Cyclotron Road, Berkeley, CA 94709\\
$^{5}$ Department of Astrophysical Sciences, Princeton University, Peyton Hall, Princeton NJ 08544, USA\\
$^{6}$ Center for Computational Astrophysics, Flatiron Institute, 162 5$^{th}$ Ave., New York, NY 10010, USA
}

\pubyear{2020}

\begin{document}
\label{firstpage}
\pagerange{\pageref{firstpage}--\pageref{lastpage}}
\maketitle

\begin{abstract}
The shapes of galaxy $N$-point correlation functions can be used as standard rulers to constrain the distance-redshift relationship and thence the expansion rate of the Universe. The cosmological density fields traced by late-time galaxy formation are initially nearly Gaussian, and hence all the cosmological information can be extracted from their 2-Point Correlation Function (2PCF) or its Fourier-space analog the power spectrum. Subsequent nonlinear evolution under gravity, as well as halo and then galaxy formation, generate higher-order correlation functions. Since the mapping of the initial to the final density field is, on large scales, invertible, it is often claimed that the information content of the initial field's power spectrum is equal to that of all the higher-order functions of the final, nonlinear field. This claim implies that reconstruction of the initial density field from the nonlinear field renders analysis of higher-order correlation functions of the latter superfluous. We here show that this claim is false when the $N$-point functions are used as \textit{standard rulers}. Constraints available from joint analysis of the galaxy power spectrum and bispectrum (Fourier-space analog of the 3-Point Correlation Function) can, in some cases, exceed those offered by the initial power spectrum even when the reconstruction is perfect. We provide a mathematical justification for this claim and also demonstrate it using a large suite of $N$-body simulations. In particular, we show that for the $z = 0$ real-space matter field in the limit of vanishing shot noise, taking modes up to $k_{\rm max} = 0.2\; h/{\rm Mpc}$, using the bispectrum alone offers a factor of two reduction in the variance on the cosmic distance scale relative to that available from the power spectrum. If these conclusions extend to the redshift-space galaxy density field, they suggest that future large-scale structure surveys may yet be able to significantly tighten dark energy constraints by design focus on higher number densities.
\end{abstract}



\section{Introduction}
In the consensus picture of cosmological structure formation, quantum fluctuations before inflation seeded cosmological density fields that were then amplified by gravitational instability (e.g. \citealt{Starobinsky_1982,Bardeen_1983}). These amplified density fields are traced by galaxies at lower redshifts, and their initial stochasticity propagates to the late-time distribution of galaxies. The statistical properties of the late-time galaxy distribution can be described by a series of $N$-Point Correlation Functions (NPCFs),
\begin{align}
    \nonumber
    \langle\delta(\textbf{r}_1)\delta(\textbf{r}_2)\rangle &\equiv \xi(\textbf{r}_1 - \textbf{r}_2),\\
    \langle\delta(\textbf{r}_1)\delta(\textbf{r}_2)\delta(\textbf{r}_3)\rangle &\equiv \zeta(\textbf{r}_1 - \textbf{r}_2, \textbf{r}_2 - \textbf{r}_3),\\
    \nonumber
    \langle\delta(\textbf{r}_1)\ldots\delta(\textbf{r}_4)\rangle &\equiv \eta(\textbf{r}_1 - \textbf{r}_2, \ldots,\textbf{r}_3 - \textbf{r}_4),\\
        \nonumber
    & \ldots,
\end{align}
where $\delta(\textbf{r})$ is the density fluctuation at position $\textbf{r}$ and the angle brackets denote ensemble averages over realizations (e.g. \citealt{1980lssu.book.....P}). Translational invariance forces the correlation functions to depend only on the difference between position vectors.\footnote{In detail, the observed galaxy density field is subject to redshift-space distortions (RSD) due to peculiar velocities, which are not translation-invariant but rather spherically symmetric about the observer (e.g. \citealt{Hamilton_1996}). However, the induced breaking of translation symmetry is small.}  For a random field that is sufficiently close to Gaussian, which is the case on large scales, a collection of all correlation functions contains all of the non-stochastic information about the field.\footnote{It has been noted that on small scales, knowing all of the zero-lag correlation functions (the moments) is not necessarily sufficient for reconstructing the one-point density PDF \citep{2011ApJ...738...86C,2012ApJ...750...28C,2017MNRAS.469.2855C}. However our goal here is not to reconstruct the one-point PDF from its cumulants, but rather to use the higher-order correlation functions as a standard ruler for the Universe's expansion.} It is often convenient to analyze the Fourier Transforms (FTs) of the NPCFs, defined as
\begin{align}
    \nonumber
    P(\mathbf{k}) & \equiv \mathcal{F}\{\xi(\mathbf{r})\},\\
    B(\mathbf{k},\mathbf{k}') & \equiv
    \mathcal{F}\{\zeta(\mathbf{r},\mathbf{r}')\},\\
    \nonumber
    T(\mathbf{k},\mathbf{k}',\mathbf{k}'') &\equiv \mathcal{F}\{\eta(\mathbf{r},\mathbf{r}',\mathbf{r}'')\},\\
    \nonumber
    & \ldots,
\end{align}
where $\mathcal{F}$ denotes a multivariate FT. The functions on the left-hand side are referred to as polyspectra. The three lowest order polyspectra are the power spectrum, the bispectrum, and the trispectrum. For the full range of scales (and wave-vectors), and in the absence of measurement systematics, binning or averaging, the information content of the polyspectra is equivalent to that of the correlation functions (see e.g.  \citealt{Hoffmann_2018} for discussion of this point for the 3PCF/bispectrum). By information content we mean the Fisher information \citep{Fisher22:MathFound,10.2307/j.ctt1bpm9r4}---the precision to which cosmological parameters can be inferred from measured NPCFs or polyspectra.

\subsection{Shapes and Rulers: Direct and Indirect}
We now present two different ways in which the polyspectra can be used for cosmology. First, at any given redshift, a polyspectrum's shape as a function of the relevant wave-vectors will encode some information about the cosmological parameters and the law of gravity. In particular, high-redshift physics such as Baryon Acoustic Oscillations (BAO), the transition between radiation and matter-domination setting the equality scale\footnote{This produces a knee in the matter transfer function at $k_{\rm eq} \sim 0.01\;h/{\rm Mpc}.$}, the contribution to the Hubble expansion by the effective energy density of relativistic species ($N_{\rm eff}$), and Silk damping setup the initial shape of the polyspectra, especially the linear-theory power spectrum. As time goes on, the nonlinear interaction of density perturbations with each other produces bispectrum, trispectrum, etc. but these are simply set by integrals of simple kernels against the linear power spectrum. These kernels come from perturbatively solving fluid equations governing the matter, and are surprisingly insensitive to the cosmological parameters.\footnote{Indeed, we use the matter-dominated form for them even despite dark energy, and it differs by less than a quarter of a percent from the matter-plus-dark energy form even at $z=0$ \citep{Hivon_1995, Bouchet_1995}. These kernels do assume GR, and modifications to gravity will alter them.} Hence, the $shape$ of the matter polyspectra is set primarily by high-redshift physics, and the ``direct'' information they contain on the cosmological parameters ultimately comes from how those parameters shape the linear power spectrum at high redshift.

However, the polyspectra can also be used indirectly, as standard rulers. In this use, they probe the overall expansion of the Universe over time, which in turn is sensitive to the cosmological parameters, as they dictate the behavior of the Friedmann equation. To use polyspectra as standard rulers, one takes the measured shape at a given redshift or redshifts and compares with the fiducial shape. This  reveals both the distance to galaxies and the expansion rate of the Universe at that time from how much the shape has dilated. This latter use of polyspectra is the focus of the present work. 

For completeness, we now briefly review previous use of the polyspectra in both of the ways outlined above. We then separately discuss density-field reconstruction.

\subsection{Direct Use of the Shape}
We first discuss previous work on direct use of the shape to constrain cosmology. The amplitude of the polyspectra at different wave-vectors is a function of the cosmological parameters. Consequently, unbiased and precise measurements of polyspectra can be used to infer the values of the cosmological parameters. For instance, the wavelength of the BAO as viewed in Fourier space (a decaying oscillatory feature in the polyspectra, e.g. \citealt{1970Ap&SS...7....3S,1970ApJ...162..815P,1998ApJ...496..605E,1999MNRAS.304..851M, Slepian_SA}) is set by the sound horizon at decoupling $(z\sim 1020)$, which in turn is  sensitive to the baryon and photon densities as well as the integrated expansion of the Universe (driven primarily by radiation and matter) of the Universe up to that point (e.g. \citealt{Hu_1995, HS_1996}). As another example, the total mass of neutrinos, $m_{\nu}$, changes the amplitude of the spectra at high wave-numbers \citep{1980PhRvL..45.1980B,1998PhRvL..80.5255H,2006PhR...429..307L,2007MNRAS.381.1313A,2009PhRvD..80h3528S,2011MNRAS.410.1647A}, and $N_\mathrm{eff}$ changes the phase \citep{2017JCAP...11..007B}. 

As earlier mentioned, the polyspectras' intrinsic shapes are set at early times and cosmological parameters that affect later-time evolution of the Universe do not imprint characteristic scales on them. For instance, small amounts of spatial curvature, $\Omega_\mathrm{K}$, parameters describing deviations from General Relativity (GR), and properties of dark energy, such as its energy density, $\Omega_\Lambda$, and the current value of its equation of state, $w_0$, do not strongly affect the polyspectras' average shape at a given redshift. 

Nonetheless, some ``direct'' constraints on these parameters can still be derived from the dependence of the polyspectra on the angle of its wave-vectors to the line of sight
thanks to Redshift Space Distortions (RSD) imprinted by velocity fields \citep{1998ASSL..231..185H,2008Natur.451..541G,2011RSPTA.369.5058P,2012MNRAS.425.2128J,2013JCAP...04..031L,2019JCAP...03..007V,2019MNRAS.485.2194H,2019JCAP...06..040W,2019MNRAS.488.1987G, Gagrani_2017, Slepian_2018, Sugi_2019, Sugi_2020, Garcia_2020, kam_2020}. 

In addition to constraining the cosmological parameters, the intrinsic shapes of galaxy polyspectra can be used to constrain bias parameters describing the connection between galaxies and their host halos \citep{2005PhRvD..71f3001S,2005MNRAS.364..620G,2007PhRvD..76h3004S,2009ApJ...702..425G,2009MNRAS.399..801G,2011MNRAS.415..383M,2011ApJ...737...97M,2011ApJ...739...85M,2013MNRAS.432.2654M, Yuan_2017, Yuan_2018}. We note that it may also be possible to extract these constraints without explicitly computing the NPCF by fitting the observed field directly \citep{2010MNRAS.407...29J,2013MNRAS.432..894J,2017JCAP...12..009S,2019MNRAS.490.4237L,2019PhRvD.100d3514S,2019JCAP...10..035H, Paco_2020b}.

\subsection{Indirect Use as Standard Rulers}
\subsubsection{Power spectrum}
Fortunately for us, the measured spectra can also be used as standard rulers (\S\ref{sec:clustering_intro}). We can compare apparent scales of different features in the spectra to the ones expected from a model to infer how far away the galaxies at various redshifts are from us \citep{2000A&A...358..395R,2010ApJ...715L.185P,2013PhR...530...87W,2018JCAP...12..014N,2019PhRvD..99l3515A,2020arXiv201011324W}. This distance-redshift relationship is very sensitive to dark energy parameters that drive the late-time time evolution of the Universe \citep{2003AAS...202.2312G,2011MNRAS.410.1993S}. This technique has become standard for the power spectrum and 2PCF, and has offered much stronger constraints on dark energy compared to what we would get from the direct dependence of the shape on the cosmological parameters \citep{2015PhRvD..92l3516A,2020arXiv200708991E}. Indeed, the dependence of the intrinsic shape of the power spectrum on dark energy parameters is very mild, and BAO constraints on dark energy in fact come from the usage of the power spectrum and 2PCF as standard rulers.

\subsubsection{Higher-order polyspectra}
Higher-order polyspectra are significantly more challenging to measure and analyze. The difficulties range from computational (the algorithms are computationally expensive, though recent work by \cite{Slepian_2015_3pt_alg, 3PCF_FFT, 3PCF_FFT_dust, Slepian_2018, Pearson_2019, Nunez_2020, Garcia_2020} has improved this for the 3PCF and anisotropic 3PCF, and augurs to do so for NPCFs as well (\citealt{Cahn_2020}, Slepian et al. in prep. 2020). A graph database approach was developed by \cite{Sabiu_2019} and improves the speed of measuring e.g the 4PCF. \cite{Tomlinson_2019} also provides a new fast scheme for measuring polyspectra, though it averages over the internal angles and tracks only the side lengths.  Algorithmic improvements have recently been made for the bispectrum \cite{Scoccimarro_2015, Pearson_2018, Sugi_2019}. Computing theory predictions is also a challenge of using polyspectra (perturbation theory does not converge nearly as well for higher orders). Higher-order polyspectra have higher dimensionality (more wave-vectors to track), which creates additional practical difficulties in computing their covariance matrices. 

The power spectrum of galaxies has been used extensively (both as a standard ruler and through its intrinsic shape) to derive high-precision cosmological constraints \citep{2005MNRAS.362..505C,2011MNRAS.416.3017B,2012MNRAS.425..405B,2012MNRAS.427.3435A,2015MNRAS.449..848H,2015MNRAS.449..835R,2015PhRvD..92l3516A,2017MNRAS.470.2617A,2018PhRvD..98d3526A,2019PhRvD..99l3505A,2020arXiv200708991E}. A similar bispectrum analysis has been attempted in recent works \citep{2001ApJ...546..652S,2015MNRAS.451..539G,2015MNRAS.452.1914G}. \citet{2017MNRAS.469.1738S} and \citet{2018MNRAS.478.4500P} used the BAO feature in the 3PCF as a standard ruler to constrain the distance-redshift relationship. \cite{Slepian_2018} proposed a convenient basis for the anisotropic 3PCF, with improvements to the line of sight in \cite{Garcia_2020}, and \citet{2019MNRAS.484..364S} proposed a similar basis for the anisotropic bispectrum and \citet{2020MNRAS.tmp.3528S} used it to measure a BAO feature in BOSS-like\footnote{BOSS is an acronym for the Baryon Oscillation Spectroscopic Survey \citep{2013AJ....145...10D}.}  mock catalogues. \citet{Slepian_3PCF_comp} measured the BOSS 3PCF in a compressed basis, and \citet{2018MNRAS.474.2109S} used the BOSS 3PCF to put the tightest current constraints on how high-redshift  baryon-dark matter relative velocities bias galaxy formation. \citet{2019MNRAS.484L..29G} and \citet{2019MNRAS.484.3713G} used the compressed BOSS bispectrum to obtain cosmological constraints.  

\cite{Sabiu_2019} measured the BOSS 4PCF using a graph-database approach, and this is the only recent work of which we are aware measuring quadruplet correlations (thee Fourier space analog of which is the trispectrum).\footnote{Our focus here is recent work, but there were a handful of papers measuring the 3PCF or bispectrum and 4PCF from the 1960s-1980s, reviewed in \cite{Peebles_2001}, beginning on page 6.} There has been theoretical work on the trispectrum (e.g. \cite{Bertolini_2016} but  it is challenging to pursue and hence compressions have recently been suggested \citep{2011MNRAS.412.1993M,2020arXiv200902290G}. Recent works have suggested that adding just the bispectrum to the standard cosmological analysis has the potential to significantly tighten derived cosmological constraints \citep{2017MNRAS.467..928G,2018MNRAS.478.1341K,2020JCAP...03..040H,2020arXiv201202200H,2020arXiv201014523M,2020JCAP...06..041G,2020arXiv200812947R,2020arXiv201100899K}.

At early times, for example, for cosmic microwave background (CMB) anisotropy data, the problem of lost information in higher orders is not as acute. Early cosmological fields are believed to be very close to being Gaussian \citep{2004PhR...402..103B,2014A&A...571A..24P,2016A&A...594A..17P,2020A&A...641A...9P,2020JCAP...07..047G}. They are fully described by their power spectrum; all higher orders are either zero or are derivable from the two-point function via Isserlis' Theorem \citep{isserlis_l_1918_1431593} (also know as Wick's Theorem). We do note that as the precision of the measurements increases, higher-order effects do become relevant even for these highly Gaussian fields, e.g. in the analysis CMB lensing by the foreground structure  \citep{1987A&A...184....1B,2006PhR...429....1L,2011JCAP...03..018L,2016PhRvD..94d3519B,2018PhRvD..98l3510B,2020A&A...641A...8P}.

\subsection{Reconstruction}
One approach to simplifying the analysis is to pursue ``reconstructing'' the initial, nearly-Gaussian, version of the observed galaxy field \citep{1999MNRAS.308..763M,2006MNRAS.365..939M,2009PhRvD..79f3523P,2009PhRvD..80l3501N}. This procedure has been successfully performed on galaxy samples and is part of the standard clustering analysis toolkit of galaxy surveys \citep{Eisenstein_2007, Pad_2009, 2012MNRAS.427.2132P,2012MNRAS.427.2146X,2014MNRAS.441.3524K}. It has been shown to make the BAO feature of the power spectrum into a much sharper standard ruler \citep{2007ApJ...664..675E}. The reconstructed field is nearly Gaussian and there is very little lost by ignoring the leftover higher-order polyspectra. Reconstruction has the additional advantage of decorrelating the measured power spectrum at each different wave-number, making the covariance matrix calculations simpler \citep{2012MNRAS.419.2949N}. 

The actual reconstruction algorithms are by no means trivial, and much recent work has focused on increasing their efficiency \citep{1997MNRAS.285..793C,2012JCAP...10..006T,2013MNRAS.432..894J,2013ApJ...772...63W,2014MNRAS.445.3152B,2015MNRAS.450.3822W,2015MNRAS.453..456B,2015PhRvD..92h3523A,2019JCAP...02..027S}. In this work, we will not be concerned with the practicalities of the approach. We will assume that by some means the perfect reconstruction has been achieved, i.e. the galaxies are at the same distance from us as they were before the reconstruction, but the field was evolved back to its Gaussian version. Even in this case the field will have some non-Gaussian features generated by non-linear biasing of galaxies \citep{1986ApJ...304...15B,1993ApJ...413..447F,1996A&A...312...11B,1999ApJ...521L..83F,1999MNRAS.308..119S}, but we will assume that on large enough scales, where galaxy bias is almost linear, these effects are subdominant. 

If nearly perfect reconstruction were achievable, one might ask if there were any point in going through a complicated analysis of higher-order statistics. The bispectrum would still be useful in constraining certain features of the cosmological model (e.g. neutrino signatures at large wave-numbers, higher-order biasing, or non-Gaussian features in the initial conditions). But would it add anything meaningful to the dark energy constraints coming from large-scale clustering \citep[see e.g. discussion in][]{2015PhRvD..92l3522S}?

After all, since the mapping from the final to the initial field is invertible on large scales, it seems obvious that no information can be lost in the process of reconstruction. And, since the initial field is fully described by its power spectrum, it is often claimed that the power spectrum of the perfectly reconstructed field would be as good of a standard ruler as $all$ of the higher-order spectra of the original nonlinear field. 

The main objective of this paper is to show that this latter claim is false. While the reconstruction conserves the information in some sense, the power spectrum of the Gaussian field is not always a better standard ruler than the combined polyspectra of the nonlinear field (\S\ref{sec:clustering_intro}). The series of nonlinear spectra can, in some circumstances, turn out to be more sensitive standard rulers and provide a stronger constraint on dark energy parameters (\S\ref{sec:quijote_argument}).

\section{Clustering as cosmological probe}
\label{sec:clustering_intro}

\subsection{Constraints from Intrinsic Shapes}

We will denote the ordered hierarchy of polyspectra by
\begin{equation}
\mathcal{S} \equiv \left[P\!\left(k^P\right), B\left(k^B_1, k^B_2\right), T(k^T_1, k^T_2, k^T_3), \ldots \right],   
\end{equation}
where the superscripts $k$ explicitly identify the parent functions of each argument (P for power spectrum, B for bispectrum, T for trispectrum). We will use a superscript $I$ for the initial (Gaussian) field and $F$ for the final (nonlinear) field.\footnote{If we require a superscript for another designation, in some contexts $I$ and $F$ will appear in the subscript.} $\mathcal{S}$ is a function of series of wave-numbers (a single wave-number for the power spectrum, pairs of wave-numbers for the bispectrum, etc.). We will denote a collection of these wave-numbers as
\begin{equation}
\label{eq:kall}
\mathbb{k} = \left[ k^P, k^B_1, k^B_2, k^T_1, k^T_2, k^T_3, \ldots \right].
\end{equation}
$S^I$ is then just the initial power spectrum, but $S^F$ goes through all higher orders. We will assume that on large scales they are both invertible as functions of the cosmological parameters $\boldsymbol{\theta}$. If that is the case, they can be uniquely mapped to each other by
\begin{equation}
\label{eq:SFtoSI}
    \mathcal{S}^F\!\!(\mathbb{k},\boldsymbol{\theta}) = \mathcal{S}^F\!\!\left(\mathcal{S}^I\!(\mathbb{k}',\boldsymbol{\theta}),\mathbb{k},\boldsymbol{\theta}\right).
\end{equation}
The above equation states that given for all cosmological parameter values, all polyspectra of the nonlinear field can be computed given the initial power spectrum.
We put $\mathbb{k}'$ as an argument of $\mathcal{S}^I$ to highlight the fact that gravitational evolution mixes modes and the final polyspectra at a specific wave-number depending on the initial power spectrum in a wide range of wave-numbers. We primed the argument to ensure that it does not get confused with a similar argument of $\mathcal{S}^F$ from which it is distinct (e.g. it does not contribute to the derivatives of $\mathcal{S}^F$ with respect to the wave-number). The $\boldsymbol{\theta}$, on the other hand, have the same values in both places they appear. 

The functional dependence (\ref{eq:SFtoSI}) is highly nonlinear and is impossible to write down in a closed analytic form. Good approximations to it can be achieved by means of perturbation theory series or $N$-body simulations. In Eulerian Standard Perturbation Theory (SPT) up to second order the equivalent of equation~(\ref{eq:SFtoSI}) is
\begin{align*}
    \nonumber
    P^F\!(\mathbf{k}) &= D^2P^I\!(\mathbf{k})\ \\
    & + 2D^4\!\displaystyle\int\!\! \left[F_2(\mathbf{k}-\mathbf{k}',\mathbf{k}')\right]^2\!P^I(\mathbf{k} - \mathbf{k}')P^I\!(\mathbf{k}')\mathrm{d}\mathbf{k}' \ \\
    \nonumber
    &  + 6D^4\!\displaystyle\int\!\! F_3(\mathbf{k},\mathbf{k}',-\mathbf{k}')P^I\!(\mathbf{k})P^I\!(\mathbf{k}')\mathrm{d}\mathbf{k}', \\
    B^F\!(\mathbf{k}_1,\mathbf{k}_2) &= 2D^4F_2(\mathbf{k}_1,\mathbf{k}_2)P^I\!(\mathbf{k}_1)P^I\!(\mathbf{k}_2) + \mathrm{cyclic}.
\end{align*}
$D$ is the linear growth factor; the form of the $F_n$ kernels is given in e.g. \citet{2002PhR...367....1B}.

In real surveys, polyspectra cannot be measured with arbitrarily fine binning in $\mathbb{k}$. The finite volume only lets us measure them at discrete sets of modes. From now on, we will write them as vectors $\mathcal{S}_i(\boldsymbol{\theta})$ where the subscript index runs over wave-number bins (or sets of wave-number bins for higher-order polyspectra) with some arbitrary ordering.

We take it that the binned final polyspectra are measured within some uncertainty described by a covariance matrix,
\begin{equation}
    \langle\delta\mathcal{S}^F_i\delta\mathcal{S}^F_j\rangle = \left[C^S_F\right]_{ij}.
\end{equation}
The subscript $F$ denotes ``final'' field and the superscript $\mathcal{S}$ (on the righthand side) that it is the covariance of a polyspectrum $\mathcal{S}$.\footnote{We will denote all covariance matrices by $\mathsf{C}$. The superscript will be used to indicate what quantity this covariance matrix is of, and the subscript will indicate the origin of the measurement. E.g. $\mathsf{C}^a_b$ indicates a covariance of $a$ when it is inferred from $b$.}
The best possible uncertainty within which the initial power spectrum can be reconstructed, and its covariance, are then given respectively by
\begin{align}
\label{eq:ftoi1}
    \delta\mathcal{S}^I_i & \sim \displaystyle\sum_j\frac{\partial\mathcal{S}^I_i}{\partial\mathcal{S}^F_j}\delta\mathcal{S}^F_j,\\
    \label{eq:ftoi2}
    \left[C^S_I\right]_{ij} &\equiv \langle\delta\mathcal{S}^I_i\delta\mathcal{S}^I_j\rangle  \sim \displaystyle\sum_{k\ell}\frac{\partial\mathcal{S}^I_i}{\partial\mathcal{S}^F_k}\frac{\partial\mathcal{S}^I_j}{\partial\mathcal{S}^F_\ell}\left[C^S_F\right]_{k\ell},
\end{align}
where $\mathsf{C}^S_I$ is a close to diagonal matrix, since the power spectrum of the initial Gaussian field is not correlated across wave-numbers. We will assume that the reconstruction procedure is perfect and the actual covariance of the reconstructed polyspectra is not significantly different from the one in equation~(\ref{eq:ftoi2}). We will also assume that the reconstruction is free of biases (e.g. the bias from assuming a wrong cosmological model, survey window effects, etc.). This issue in the context of BAO reconstruction is discussed in \cite{Sherwin_2019}.

The best precisions one can get for the cosmological parameters from, respectively, the initial and final polyspectra are
\begin{align}
\label{eq:itop}
    \left[C^\theta_I\right]^{-1}_{k\ell} &=\displaystyle\sum_{ij}\frac{\partial\mathcal{S}^I_i}{\partial \theta_k} \frac{\partial\mathcal{S}^I_j}{\partial \theta_\ell}\left[C^S_I\right]^{-1}_{ij},\\
    \label{eq:ftop}
    \left[C^\theta_F\right]^{-1}_{k\ell} &=\displaystyle\sum_{ij}\frac{\partial\mathcal{S}^F_i}{\partial \theta_k} \frac{\partial\mathcal{S}^F_j}{\partial \theta_\ell}\left[C^S_F\right]^{-1}_{ij}.
\end{align}
To compare these precisions we need to relate the derivatives of the initial and the final polyspectra with respect to cosmological parameters (i.e. the sensitivity to cosmology) bin by bin. From equation~(\ref{eq:SFtoSI}) this relationship is
\begin{equation}
\label{eq:full_derivatives}
\frac{d\mathcal{S}^F_i}{d\boldsymbol{\theta}} = \displaystyle\sum_j\frac{\partial\mathcal{S}^F_i}{\partial\mathcal{S}^I_j}\frac{\partial\mathcal{S}^I_j}{\partial\boldsymbol{\theta}} + \frac{\partial\mathcal{S}^F_i}{\partial\boldsymbol{\theta}}.
\end{equation}
The second term in this expression describes how much the mapping between initial and final spectra itself depends on cosmological parameters. In cosmological models without dark energy this derivative is negligible; to compute evolved polyspectra at all wave-numbers, one only needs to know the linear power spectrum but not the value of $\Omega_\mathrm{m}$ \citep{1991ApJ...371....1M,1994ApJ...433....1B}. In the presence of a time-independent cosmological constant this is not formally true, but the actual dependence of the mapping on $\Omega_\mathrm{m}$ and $\Omega_\Lambda$ is extremely weak \citep{1998ApJ...496..586S,1998MNRAS.301..535F}. This also seems to be the case for time-dependent dark energy models on large scales \citep{2006MNRAS.366..547M}.

Hence, on large scales and for conventional cosmological models, the above equation simplifies to
\begin{equation}
\label{eq:pderivatives}
\frac{d\mathcal{S}^F_i}{d\boldsymbol{\theta}} = \displaystyle\sum_j\frac{\partial\mathcal{S}^F_i}{\partial\mathcal{S}^I_j}\frac{\partial\mathcal{S}^I_j}{\partial\boldsymbol{\theta}}.
\end{equation}
Combining this result with equations~(\ref{eq:ftoi1}) and (\ref{eq:ftoi2}) ensures that the cosmological constraints derivable from the initial and final polyspectra given by $C^{\boldsymbol{\theta}}_I$ and $C^{\boldsymbol{\theta}}_F$ are exactly equal. This is the sense in which the information is conserved by the reconstruction. The origin of this conservation is also clear. The covariance and the sensitivity of the polyspectra scale as inverses of each other under reconstruction. This conforms with our intuition that invertible transformations should conserve information. 

It is worth reflecting on what equations~(\ref{eq:itop}) and (\ref{eq:ftop}) actually mean. They both tell us that if we were given a box with a galaxy field, from which we measure polyspectra with some accuracy, the cosmological constraints derivable from those polyspectra depend on two things: how well one measures each spectrum at different wave-numbers, and how sensitive that wave-number is to the cosmological parameters (i.e. by how much would it change if we changed $\boldsymbol{\theta}$ by a small amount). The procedure is that we are fitting to each mode independently (except that the cross-correlation in measurements is accounted for in the covariance matrices) and are summing up the constraint coming from all the modes of all the polyspectra. These equations are appropriate when the fits are performed on the intrinsic shape of the polyspectra (e.g. constraining the amplitude and amplitude-like signals such as that of the neutrino mass). However, they are not appropriate when the polyspectra are used as standard rulers.

To see why this is the case, imagine a hypothetical universe in which the polyspectra do not depend on cosmological parameters, perhaps due to some accidental cancellation of linear and nonlinear evolution effects. No matter what set of cosmological parameters we consider, we always have the same initial conditions and the same nonlinear evolution. Equations~(\ref{eq:ftoi1}) and (\ref{eq:ftoi2}) would then tell us that we cannot extract constraints on $\boldsymbol{\theta}$ from measured polyspectra, since the derivatives are zero. 

This result makes perfect sense. Given a box of galaxies, we cannot guess what the input cosmological parameters were if all cosmological parameters result in the polyspectra of identical shapes. It is also clear that for real observations, we could still use these polyspectra as standard rulers, by comparing their known sizes in physical units to their apparent sizes on the sky. We can then use these measurements of distance to constrain dark energy (as long as the expansion history of the Universe still depends on dark energy). In a certain sense, if the intrinsic shapes of the polyspectra did not depend on cosmological parameters they would make even better standard rulers, because we would not have to worry about cosmology dependence of the template. This toy model suggests that when computing the constraining power of a standard ruler we need to modify our formalism. The geometrical probes contain additional information to the one that can be retrieved just from looking at the intrinsic shape of NPCF.

\subsection{Constraints from Standard Rulers}

We will start this section with a brief recap of how the standard ruler measurements with galaxy polyspectra are made. Appendix~\ref{app:standardruler} provides a more detailed review. Since we observe galaxies on the sky and do not know how to convert distances and angles to physical separations $a$ $priori$, we initially compute polyspectra in some fiducial cosmology. Unless we are lucky, the fiducial cosmology will not be identical to the real cosmology of the Universe. The measured power spectrum  $P_\mathrm{m}(k)$ will have the same intrinsic shape as the true power spectrum, $P_\mathrm{t}(k)$, but since we did not put galaxies at correct distances, it will be uniformly dilated with respect to the real power spectrum in physical units as
\begin{equation}
P_\mathrm{m}(k) = P_\mathrm{t}(\alpha k).
\end{equation}
\noindent
The dilation parameter $\alpha$ is the ratio of the fiducial distance we assumed to the true distance to the galaxies: $\alpha = D_\mathrm{t}(z)/D_\mathrm{fid}(z)$, where $D$ is the comoving distance (see Appendix~\ref{app:standardruler} for details). By measuring it from the power spectrum, we can estimate the distance to a galaxy sample at redshift $z$. This distance-redshift relationship depends on cosmological parameters (including dark energy parameters) and can be used to obtain very accurate cosmological constraints even if the intrinsic shape of the power spectrum does not have such a dependence.

Small uncertainties in the measured power spectrum propagate to small uncertainties in the recovered value of $\alpha$ via
\begin{equation}
\label{eq:dPdalpha}
    \delta P_\mathrm{m}(k) \sim \frac{\partial P_\mathrm{m}(k)}{\partial \alpha}\delta\alpha = \frac{\partial{P_\mathrm{t}(\alpha k})}{\partial\ln k}\delta\alpha.
\end{equation}
This equation tells us that the power spectrum makes a better standard ruler if its intrinsic shape changes rapidly in wave-number. If the power were constant across wave-numbers, we would not be able to use it as a standard ruler. The presence of characteristic scales and oscillations (e. g. the BAO) where the slope rapidly changes with the wave-number makes the power spectrum a better standard ruler than it would have been if it were a smoother shape. How sensitive the intrinsic shape is to the variations in cosmological parameters is, in this case, irrelevant. The intrinsic shape could have a very weak or no dependence on cosmological parameters for all we care. The important thing is how sensitive is the spectrum to uniform dilation (given by equation~\ref{eq:dPdalpha}) and how sensitive is the distance-redshift relationship to the change in cosmology ($\partial\alpha/\partial\theta$). 

The actual standard ruler analysis is much more complicated than this. There are two $\alpha$ parameters describing the line-of-sight and across-the-line-of-sight dilation and we have to properly account for the fact that the cosmology dependence is both in the dilation (standard ruler effect) and in the template (intrinsic dependence of the shape). None of this affects the essence of our main argument, which is that the goodness of the field as a standard ruler is not conserved under gravitational evolution. For simplicity, we will assume that the standard ruler is used to constrain a single scale.

Use of the bispectrum as a standard ruler is similar. We relate measured bispectra in the fiducial cosmology to a template via
\begin{equation}
    B_\mathrm{m}(k_1,k_2) = B_\mathrm{t}(\alpha k_1,\alpha k_2),
\end{equation}
\noindent
and how good of a standard ruler the bispectrum is will be determined by
\begin{equation}
\label{eq:dBdalpha}
    \delta B_\mathrm{m}(k_1,k_2) \sim \left[\frac{\partial B_\mathrm{t}(k_1,k_2)}{\partial \ln k_1} + \frac{\partial B_\mathrm{t}(k_1,k_2)}{\partial \ln k_2}\right]\delta\alpha.
\end{equation}

The precision to which $\alpha$ can be measured from the initial and the final polyspectra is
\begin{align}
   \label{eq:itoa}
    \left[C^\alpha_I\right]^{-1} &= \displaystyle\sum_{ij}\frac{\partial\mathcal{S}^I_i}{\partial \alpha} \frac{\partial\mathcal{S}^I_j}{\partial \alpha}\left[C^S_I\right]^{-1}_{ij} =\displaystyle\sum_{ij}\frac{\partial\mathcal{S}^I_i}{\partial \ln\mathbb{k}} \frac{\partial\mathcal{S}^I_j}{\partial \ln\mathbb{k}}\left[C^S_I\right]^{-1}_{ij},\\
    \nonumber
    \left[C^\alpha_F\right]^{-1} &= \displaystyle\sum_{ij}\frac{\partial\mathcal{S}^F_i}{\partial \alpha} \frac{\partial\mathcal{S}^F_j}{\partial \alpha}\left[C^S_F\right]^{-1}_{ij} =\displaystyle\sum_{ij}\frac{\partial\mathcal{S}^F_i}{\partial \ln\mathbb{k}} \frac{\partial\mathcal{S}^F_j}{\partial \ln\mathbb{k}}\left[C^S_F\right]^{-1}_{ij},
\end{align}
where the derivatives with respect to $\mathbb{k}$ denote a combined derivative of the polyspecrtra with respect to all wave-numbers as in 
\begin{equation}
    \frac{\partial{\mathcal{S}}}{\partial \ln \mathbb{k}} \equiv \displaystyle\sum_i\frac{\partial\mathcal{S}}{\partial \ln \mathbb{k}_i},
\end{equation}
where the components of $\mathbb{k}$ are given by equation~(\ref{eq:kall}).
This sum will contain the terms given by equations~(\ref{eq:dPdalpha}) and (\ref{eq:dBdalpha}) and similar derivatives for the higher-order NPCFs.

The precision to which cosmological parameters can be derived from $\alpha$ is \begin{align}
   \label{eq:atop}
    \left[C^{\boldsymbol{\theta}}_\alpha\right]^{-1}_{ij} &=\frac{\partial\alpha}{\partial\boldsymbol{\theta}_i} \frac{\partial\alpha}{\partial\boldsymbol{\theta}_j}\left[C^\alpha\right]^{-1},
\end{align}
$\mathsf{C}^\alpha$ is the covariance matrix of $\alpha$ and can be either $\mathsf{C}^\alpha_I$ or $\mathsf{C}^\alpha_F$ and since $\alpha$ is a single number this is just its variance.
The equation above does not change between the initial and the final fields. How the $\alpha$ errors propagate to the cosmological parameter errors depends on the variance of $\alpha$ but does not depend on the origin of the measurement. To determine which standard ruler (initial or final) results in better constraints we have to compare the first and second equations of~(\ref{eq:itoa}).

The covariance matrices scale by equation~(\ref{eq:ftoi2}) as before. The shape derivatives of the final polyspectra, on the other hand, are
\begin{equation}
\label{eq:kderivative}
    \frac{\partial \mathcal{S}^F}{\partial\ln\mathbb{k}} = \frac{\partial}{\partial\ln\mathbb{k}}\left\lbrace\mathcal{S}^F\!\!\left(\mathcal{S}^I\!(\mathbb{k}',\boldsymbol{\theta}),\mathbb{k},\boldsymbol{\theta}\right)\right\rbrace.
\end{equation}
Unlike equation~(\ref{eq:pderivatives}), where we were able to claim a simple relationship between the two derivatives, there is no general relationship that relates the above to $\partial\mathcal{S}^I/\partial\ln\mathbb{k}$ in this case. The difference in the two covariance matrices in equation~(\ref{eq:itoa}) does not get canceled by the difference in the derivatives.

This would be the case even if the mapping between initial and final polyspectra were linear, for instance as in
\begin{equation}
    \mathcal{S}^F\!\!(\mathbb{k},\boldsymbol{\theta}) = \displaystyle\int\!\!\! K(\mathbb{k},\mathbb{k}',\boldsymbol{\theta})\mathcal{S}^I\!(\mathbb{k}',\boldsymbol{\theta}) \mathrm{d}\mathbb{k}'.
\end{equation}
The only kernel $K$ that can enforce the equality $\mathsf{C}^\alpha_I = \mathsf{C}^\alpha_F$ is a wave-number independent scaling $K(\mathbb{k},\mathbb{k}',\boldsymbol{\theta}) \propto \delta^\mathrm{D}(\mathbb{k} - \mathbb{k}')$ where $\delta^\mathrm{D}$ is a Dirac delta function.

On reflection, this is not surprising. The cosmological evolution does not know that we will be using polyspectra as standard rulers, so barring a coincidence there is no reason why it should conserve the strength of the standard rulers over time.

To summarize, even though the evolution from the initial to the final field is invertible on large scales, it does not conserve the field's cumulative capacity as a standard ruler. How good of a standard ruler the polyspectra are is determined by how sensitive they are with respect to wave-number dilation, or how rapidly they change across wave-numbers. This property is not conserved by gravitational evolution simply because there is no underlying mathematical reason for it to be conserved.

\section{Tests on Simulations}
\label{sec:quijote_argument} 

In the previous section, we established that the power spectrum of the initial Gaussian field does not have to provide the same constraints on the scale parameter $\alpha$ as the hierarchy of polyspectra of the final nonlinear field. Which one is a better standard ruler is difficult to guess without performing actual calculations. The explicit analytic calculations, unfortunately, are difficult to make. Doing so would require explicitly writing down the mapping given by equation~(\ref{eq:SFtoSI}) between the polyspectra of the initial and the final fields. 

The power spectrum of the nonlinear field is a worse standard ruler than the power spectrum of the Gaussian field for two reasons. Gravitational evolution partially erases the BAO feature making the nonlinear power spectrum smoother (smaller derivative in equation~\ref{eq:dPdalpha}). The nonlinear power spectrum is also more correlated between different wave-numbers, making it a less powerful measurement overall. On the other hand, the nonlinear field gains additional polyspectra, like a bispectrum, that can be used as standard rulers. The real question is whether the  bispectrum as a standard ruler is good enough to make up for the lost constraining power due to partially erased BAO wiggles in the nonlinear power spectrum.

There is no universal answer to this question. For specific galaxy samples the answer depends on many factors such as the number density, galaxy bias, range of wave-numbers used in fitting, fitting procedure, etc. Our goals is not to argue that the nonlinear polyspectra are better standard rulers but to to demonstrate that there is no fundamental reason why they cannot be.

To make our results as transparent as possible, we take a real-space matter field at redshift zero. Performing the same analysis in redshift-space would significantly complicate it (introduction of two dilation parameters, measuring anisotropic bispectrum), and for what we are trying to show it would not matter. If anything, we expect our results to be even stronger in redshift-space, where the bispectrum has a more non-trivial shape.

One of the difficulties of analyzing higher-order polyspectra is the need for accurate covariance matrices. Measuring the bispectrum up to $k_\mathrm{max} = 0.2h\ \mathrm{Mpc}^{-1}$ in bins of $\Delta k = 0.01h\ \mathrm{Mpc}^{-1}$ results in $925$ bispectrum measurements. An accurate estimate of the $925\times925$ covariance matrix requires at least a few thousand simulations.

In our tests we used 8,000 \texttt{Quijote} simulations \citep{Quijote} of $1h^{-3}\ \mathrm{Gpc}^3$ volume each. We compute the angle-averaged power spectrum and bispectrum in bins of $\Delta k = 0.1h\ \mathrm{Mpc}^{-1}$. We estimate the covariance of the measurements from the sample variance between 8,000 simulations (see Appendix~\ref{app:technical_details} for the details of these computations). We use equations~(\ref{eq:itop}) and (\ref{eq:ftop}) to estimate the information on the parameters coming from the intrinsic shape, and equations~(\ref{eq:itoa}) to estimate the information coming from their usage as standard rulers. Appendix~\ref{app:technical_details} explains the details of how these derivatives are computed.

\begin{figure}
	\includegraphics[width=\columnwidth]{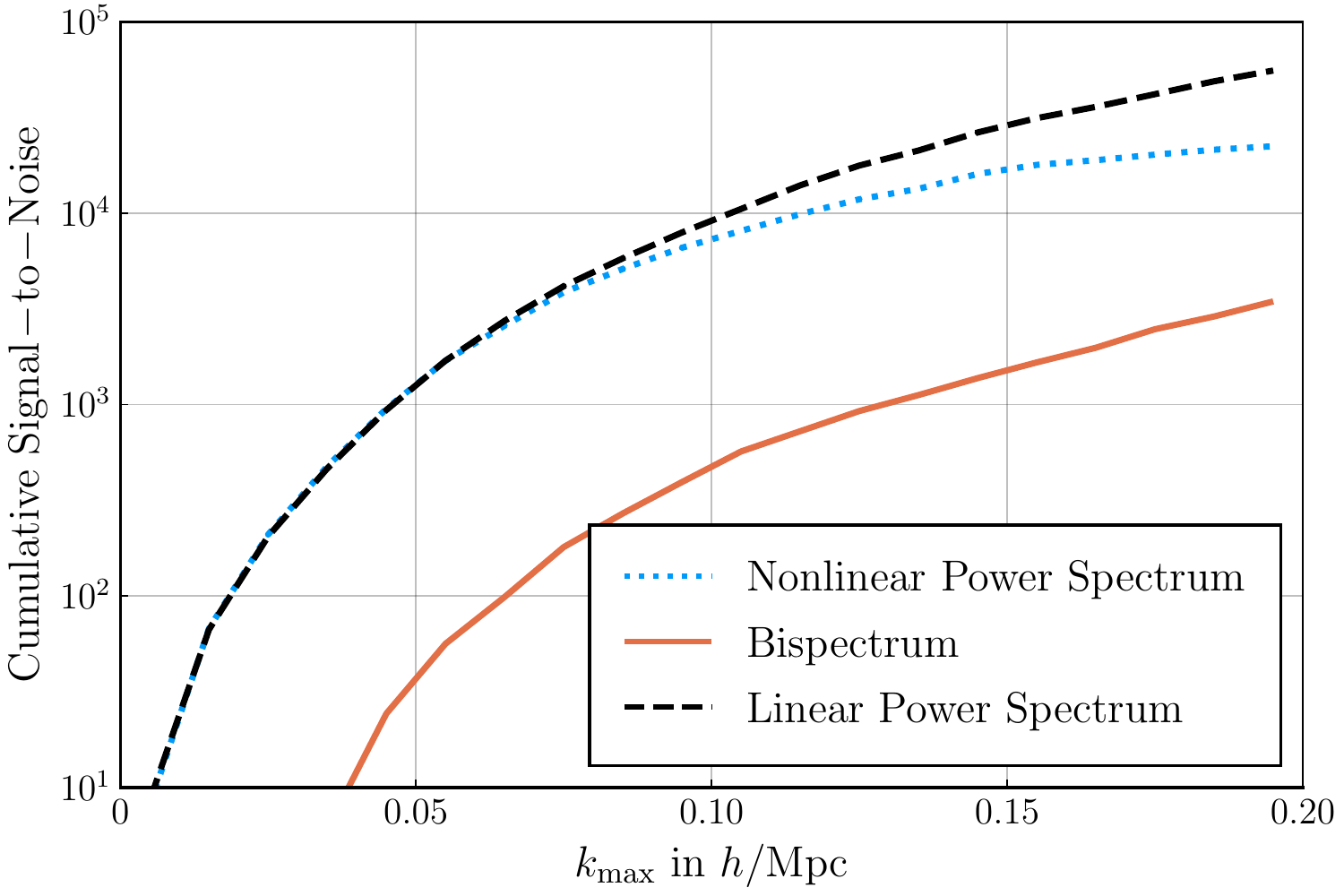}
    \caption{Cumulative signal-to-noise ratio of respectively the Gaussian power spectrum (black line), the nonlinear power spectrum (blue line), and the nonlinear bispectrum (orange line), as a function of maximum wave-number considered in the analysis. This comes from the direct shape and therefore the linear power spectrum contains all the information. Derived for the  $1h^{-3}\ \mathrm{Gpc}^3$ box of matter distribution (negligible shot-noise) at $z=0$.}
    \label{fig:SNinfo}
\end{figure}

Fig.~\ref{fig:SNinfo} shows the cumulative signal-to-noise ratios of the power spectrum and bispectrum measured from $\texttt{Quijote}$ boxes,
\begin{equation}
    \mathrm{S/N} = \displaystyle\sum_{ij} \left[\mathbf{PB}\right]_i \left[C^{-1}\right]_{ij} \left[\mathbf{PB}\right]_j^\mathrm{T},
\end{equation}
where $\mathbf{PB}$ is a row vector constructed from the power spectrum and bispectrum bins and $\mathsf{C}$ is their covariance matrix. For each point on the horizontal axis, we used all the power spectrum bins up to that value of $k_\mathrm{max}$ and all the bispectrum bins that have all their wave-numbers below that $k_\mathrm{max}$. 

We compute the cumulative signal-to-noise because it seems to be a very popular proxy for the information content of the spectra in the recent literature. In terms of parameter fits, it corresponds to the case where the power spectrum (bispectrum) intrinsic shape is known exactly and the amplitude needs to be determined from data, i.e. fitting to unknown $A_\mathrm{P}$ and $A_\mathrm{B}$ in $A_\mathrm{P}P(k)$ and $A_\mathrm{B}B(k,k',k'')$. On large scales, the measurement of $A_\mathrm{P}$ then equivalent to the measurement of $(b_1\sigma_8)^2$ and $A_\mathrm{B}$ to that of $(b_1\sigma_8)^4$, where $b_1$ is the linear bias and $\sigma_8$ (or $\sigma_{12}$) is one of the possible parametrizations of the amplitude of matter clustering \citep{2010ASPC..426..158F,2020PhRvD.102l3511S}. 

The lines on Fig.~\ref{fig:SNinfo} show signal-to-noise as a function of the maximum wave-number considered in the analysis for the nonlinear power spectrum, nonlinear bispectrum, and linear power spectrum perfectly reconstructed and placed at redshift zero. As expected, the bispectrum contains significantly less information on the amplitude and this information grows with the maximum wave-number. Yet we notice that the information extracted from the joint fit is always below the information extractable from the linear power spectrum. This result is not surprising. The amplitude is measured in the intrinsic shape of the polyspectra bin by bin, and the results are in line with our expectation from equations~(\ref{eq:itop}) and (\ref{eq:ftop}).

\begin{figure*}
	\includegraphics[width=\textwidth]{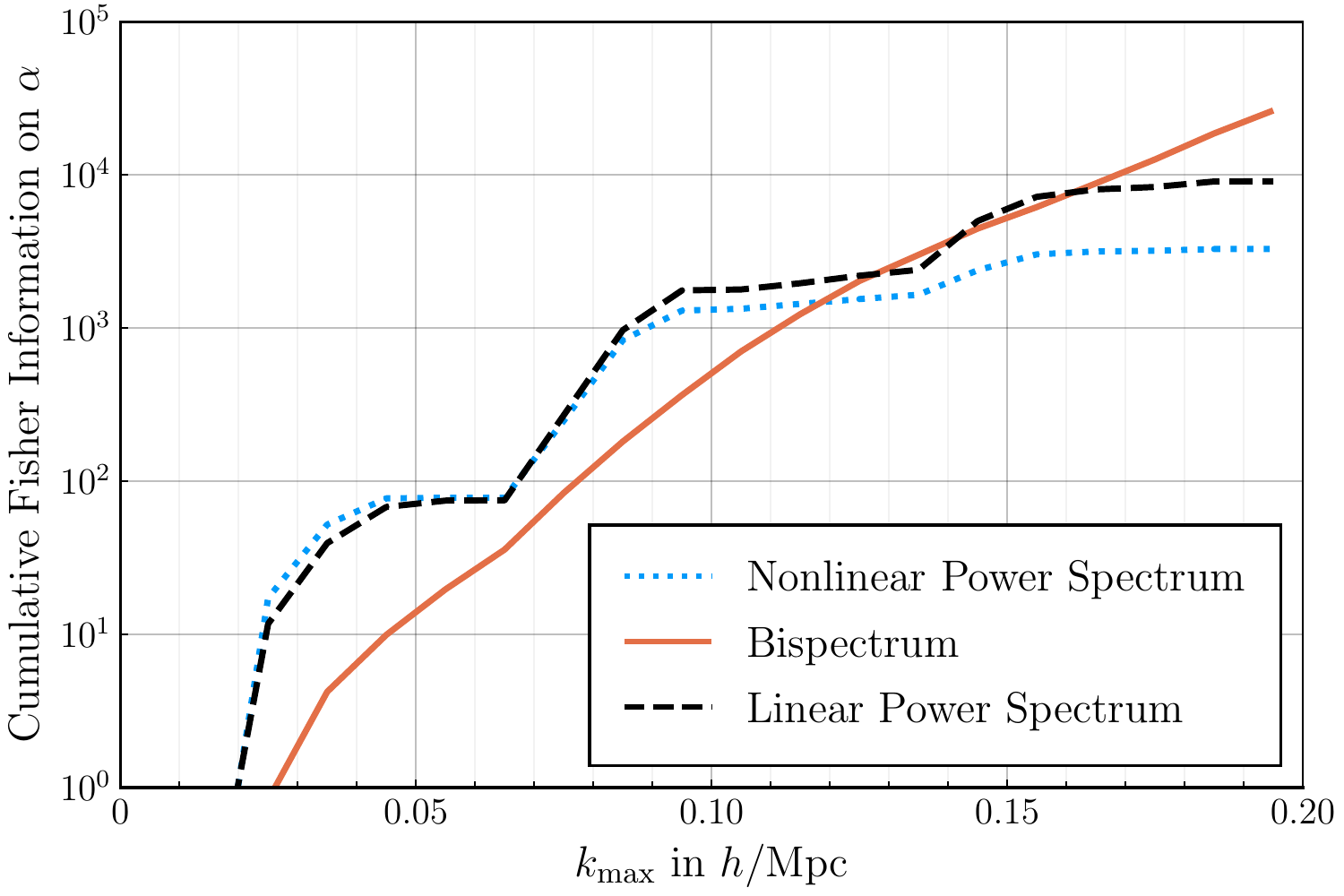}
    \caption{The information on the isotropic dilation parameter obtainable from the Gaussian power spectrum (black line), the nonlinear power spectrum (blue line), and the nonlinear bispectrum (orange line), as a function of maximum wave-number considered in the analysis. The constraints from the nonlinear field (bispectrum) exceed the ones obtainable from the initial field (linear power spectrum). Derived for the $1h^{-3}\ \mathrm{Gpc}^3$ box of matter distribution (negligible shot-noise) at $z=0$. The Fisher information of the power spectrum flattens around the peaks of the BAO features. A region remains flat until you go over the BAO peak. Three flat regions correspond to the three BAO peaks traversed (see Fig.~\ref{fig:Pk}). This is not as apparent in the bispectrum since at each $k_1$ we sum over multiple $k_2$ and $k_3$.}
    \label{fig:APinfo}
\end{figure*}

Fig.~\ref{fig:APinfo} shows a similar plot of the information on the dilation parameter $\alpha$. These constraints come from the usage of the polyspectra as standard rulers (not the intrinsic shape) and are described by equation~(\ref{eq:itoa}). These equations do not force the cumulative information in the final and initial fields to be equal, and it is indeed the case that at about $k = 0.13h\ \mathrm{Mpc}^{-1}$ the bispectrum becomes a better standard ruler for constraining $\alpha$ then the initial power spectrum. By $k_\mathrm{max} = 0.2h\ \mathrm{Mpc}^{-1}$ the improvement reaches a factor of two. 

The step-like structure with increasing $k_{\rm max}$ in the two lines corresponding to the power spectrum information is not a numerical artifact. These steps appear because the main feature in the power spectrum---the BAO wiggles---are stronger standard rulers at the edges of a given wave-form (more sensitive to dilation) than at their minima and maxima. The slope of the information curve is therefore flattened as we pass over peaks in the BAO wiggles. The low-redshift power spectrum line is slightly above the Gaussian power spectrum at very low wave-numbers for a similar reason: the nonlinear evolution moves the crest of the BAO to slightly higher wave-numbers. Technical details behind these computations are presented in Appendix~\ref{app:technical_details}.

One could argue that comparing the Gaussian power spectrum and the bispectrum at a fixed $k_\mathrm{max}$ is not fair since the bispectrum ``siphons the information'' from smaller scales in the initial field. This is certainly true, but our findings suggest that the bispectrum is a much better standard ruler at $k_\mathrm{max} = 0.2h\ \mathrm{Mpc}^{-1}$ even when compared to the Gaussian power spectrum up to $k_\mathrm{max} = 0.5h\ \mathrm{Mpc}^{-1}$. 

\section{Conclusions}
\label{sec:conclusions}
In the previous sections we showed that, in general, there is no relationship between how good standard rulers from the nonlinear field are compared to their Gaussian field counterpart. Whether the reconstructed power spectrum or the nonlinear bispectrum is a better standard ruler will depend on the specifics of a galaxy sample such as its redshift, number density and bias \citep{2017MNRAS.467..928G,2017PhRvD..96b3528C,2019MNRAS.483.2078Y,2019MNRAS.482.4883C,2019MNRAS.490.5931P,2020JCAP...03..040H,2020JCAP...06..041G,2020MNRAS.497.1684S,2020arXiv200810199L,2020arXiv201202200H}. It is undeniable that the linear power spectrum is significantly easier to analyze due to the size of the data, and the ease of modeling and computing covariance matrices. Even if the bispectrum is a better standard ruler in principle, we may not be able to reliably extract this information in practice. 

On the other hand, our tests on \texttt{Quijote} simulations suggest that at redshift zero the real-space nonlinear matter field is a better standard ruler by a factor of two! There is no reason why going to the redshift-space will reverse this order of precedence. If anything, the bispectrum analysis should benefit more from the addition RSD. We presented our main results for the bare-bones case of the real-space unbiased tracers to keep the physical picture simple, but we checked that adding extra nuisance parameters accounting for e.g. bias parameters and non-Poissonian shot-noise does not affect our conclusions.

One may wonder why it is that for the BOSS \citep{2013AJ....145...10D} and eBOSS \citep{Dawson_2016} samples the constraints coming from the joint analysis of the power spectrum and the bispectrum are slightly lower than the constraints from the reconstructed power spectrum. We think this is very likely due to the effect of shot-noise in those samples. The shot-noise affects the variance of the high wave-number modes more than it affects that of the low wave-number modes. The bispectrum starts overtaking the linear power spectrum as a standard ruler at wave-numbers of $k\sim 0.15h\ \mathrm{Mpc}^{-1}$. As the shot-noise increases, the contribution of these wave-number bins gets down-weighted. This and the fact that the nonlinear bias terms are expected to affect the bispectrum more at higher wave-numbers may be the reason behind the apparent ``conservation of information'' between nonlinear and linear fields in the BOSS and eBOSS samples.

The Bright Galaxy Survey (BGS) sample at low redshifts and the Emission Line Galaxy (ELG) sample at around $z \sim 1$ from the Dark Energy Spectroscopic Instrument (DESI; 2019-2024) experiment, will have a significantly higher number density. We expect for these samples that the higher-order polyspectra will perform especially well. On a longer time horizon, the ATLAS (Astrophysics Telescope for Large Area Spectroscopy) probe \citep{2019arXiv190900070W} is designed to provide a galaxy sample with a density of $n = 10^{-2}h^3\ \mathrm{Mpc}^{-3}$, and the forecasts for the higher-order analysis look very promising \cite{2019BAAS...51c.508W}. 21-cm intensity mapping surveys \citep{2012RPPh...75h6901P,2014JCAP...09..050V,2015ApJ...803...21B}, which have very low shot-noise, are another potentially rich candidate for use of these methods \citep{2006MNRAS.366..213S,2015MNRAS.451..266Y,2018MNRAS.476.4007M,2020MNRAS.493..594B}. Detailed forecasts for these surveys are very complicated and fall outside of the scope of this paper.

Another interesting question is what happens for the NPCFs of orders four and higher. This question is very difficult to answer without performing specific calculations. We were unable to derive reliable estimates for the trispectrum from the \textit{Quijote} simulations at high enough wave-numbers. The number of simulations, even though large for other purposes, was not large enough to reliably compute covariance matrices for the large number of trispectrum configurations.\footnote{\citet{2020arXiv200902290G} presented integrated trispectrum and its covariance estimates from 5,000 simulations, but their measurements extended only up to $k_\mathrm{max} = 0.12h\ \mathrm{Mpc}^{-1}$. The number of distinct trispectra scales as $k_\mathrm{max}^3$ and going to $k_\mathrm{max} = 0.2h\ \mathrm{Mpc}^{-1}$ with roughly the same accuracy would require up to an order of magnitude more simulations.} In general, the higher-order polyspectra do provide additional standard rulers but are also significantly noisier. It may well be that the bispectrum provides a sweet spot where the increased noise is compensated by the additional sensitivity to dilation, and for higher orders the noise scale too steeply for them to make a reasonable contribution. There is no reason why if the bispectrum is a better standard ruler than the power spectrum, the trispectrum has to be an even better standard ruler. Such an intuition would be borrowed from perturbation theory, where if a certain infinite series diverges for lower orders it must also diverge for higher orders. But the problem at hand has nothing to do with the perturbative expansion in the linear field, and so we do not think this intuition is necessarily applicable here.

There are a few reasons why the conclusions of this paper may at first seem counter-intuitive. One of them is due to the coincidence noted at the beginning of the section---the BOSS and eBOSS samples just happened to have number densities and biases that resulted in apparent conservation of information under reconstruction. Another source of this uneasiness is the basic intuition from statistics telling us that invertible transformations cannot create or destroy information. If one made certain measurements and recorded proper covariances, one can multiply these measurements by some numbers, raise them to a power, or apply a wide range of nonlinear transformations. As long as those transformations are invertible\footnote{Data reduction techniques such as binning are not always invertible.} and properly accounted for in the covariance matrices, they are not going to affect the amount of information one can extract from the fields. Current literature tends to use the signal-to-noise as a universal proxy for the information content, and for the signal-to-noise and other amplitude-like parameters the information is indeed conserved. 

In \S\ref{sec:clustering_intro} we explained why this intuition fails with standard rulers. The standard ruler tests rely on apparent effects that do not really exist in nature as such. One would struggle to identify the parts referring or related to a standard ruler test in the Euler or the Poisson equations. Therefore there is no fundamental reason why the laws of nature should conserve the efficiency of standard rulers as cosmological probes over time. In fact, the sensitivity of polyspectra to cosmological parameters sometimes is not conserved even when the constraints are coming from the analysis of their intrinsic shape (as opposed to their usage as standard rulers). To derive ``the conservation of information'' we had to assume that the mapping between the initial and the final polyspectra depends on the cosmological parameters weakly in equation~(\ref{eq:full_derivatives}). This is not the case e.g. when constraining the mass of neutrinos. In this case the mapping depends on the value of the parameter (given exactly the same initial power spectrum, the damping of the amplitude at high wave-numbers depends on the total mass of neutrinos). The evolution itself imprints a useful feature into the polyspectra and the reconstructed version would obviously have much lower sensitivity to the parameter of interest.

We can think of a few toy models to put our intuition at ease. One hypothetical example is of a universe that has a flat power spectrum initially. Let us suppose that the laws of gravity in this universe are such that they start imprinting a ``hump'' in this initially flat power spectrum with time, so that at later times we have a power spectrum with a feature. At late times then we have a standard ruler that can be used to measure the distance-redshift relationship and derive dark energy constraints. If we reconstruct the field we will go back to the featureless power spectrum that cannot be used as a standard ruler. In this hypothetical universe the gravitational evolution actually creates ``information'' out of nowhere, the information is clearly not conserved, and reconstruction would erase this information.

An even simpler example is an initial field that has only one wave-number and a single phase, and creates a sinusoidal pattern across the sky. The initial power spectrum is in this case a Dirac delta function of the special frequency. This special frequency would be much more easily detected in the nonlinearly-evolved field.

A less hypothetical example is that of standard candles. The standard candles are in many respects similar to standard rulers. They have a known intrinsic luminosity (or a luminosity that $can$ be standardized) and by measuring their apparent luminosity we can put very stringent constraints on the distance-redshift relationship and consequently on dark energy. These standard candles are used as individual objects. In standard analyses we do not care how they are arranged in space with respect to each other. These objects were also created from initial Gaussian fields with tiny fluctuations that were fully described by their power spectrum. One could ask a similar question of them: where does the information on the dark energy they provide at late times come from? Which part of the initial Gaussian power spectrum is it encoded in? It is clear that this information does not really come from anywhere. It does not reside in the small patch of the Universe that later collapsed to make a supernova, and it clearly does not reside in the primordial power spectrum. We were lucky that the standard candles happened to exist in the Universe and cosmologists had sufficient ingenuity to realise they could be used to measure distances. We were also lucky that the distance as a function of redshift happened to depend very strongly on dark energy. 

The situation with polyspectra and distance measurements is similar. Low-redshift polyspectra can be used as standard rulers. They also happened to be generated from the seed Gaussian fields. But their origin is largely irrelevant for their usage as standard rulers. When a carpenter uses a meter stick to measure the width of a window, the information about the width of that window does not come from the wood that made the meter stick. The information comes the carpenter's use of the meter stick in a certain way. The carpenter can measure as many lengths and widths as he or she wishes. There is no limit on how much information he or she can collect, at least no limit imposed by the physical origin of the meter stick.

Another reason why we feel the information must come from the Gaussian power spectrum is that we often choose to represent the information content of a random fields by the signal-to-noise ratio. For the signal-to-noise and other amplitude-like parameters, equations~(\ref{eq:itop}) and (\ref{eq:ftop}) hold. Fig.~\ref{fig:SNinfo} offers a simple and familiar example of the amplitude-like parameters for which the information is conserved, although it is not clear what kind of information it is exactly. For uncorrelated data, signal-to-noise is the sum of all measurements divided by their errors.

If one makes independent observations of the same thing many times over, the signal-to-noise is a measure of how well that quantity can be constrained cumulatively by all the measurements. When the data are measurements of different things, the signal-to-noise does not really have a clear meaning. The power spectrum measurements at two different wave-numbers are a measurement of two different physical quantities. The dependence of the two on the underlying parameters is different. It is therefore not entirely clear what the signal-to-noise represents in this case. 

The signal-to-noise rises as the quality of our measurements increases so it can be used as a reasonably good proxy for the relative goodness of two pieces of data, but in some cases this kind of comparison can be misleading. E.g. while samples with higher signal-to-noise result in better BAO constraints, they also result in worse RSD constraints \citep{2016MNRAS.463.2708P}. Computing the signal-to-noise of the power spectrum and the bispectrum bin by bin and summing it up only makes sense if one is interested in how well the amplitude of the polyspectra would be measured if their shapes were perfectly known. These kind of fits are rarely performed in practice.

\section*{Acknowledgements}
We would like to thank David Spergel and Ben Wandelt for helpful comments on the manuscript. LS and ZS are grateful to the Munich Institute for Astro and Particle Physics (MIAPP) of the DFG cluster of excellence ``Origin and Structure of the Universe'' for the invitation to the ``Dynamics of Large-Scale Structure Formation - 2019'' program where the initial discussions leading to this work were initiated. LS is grateful for support from DOE grants DE-SC0021165 and DE-SC0011840, NASA ROSES grants 12-EUCLID12-0004 and 15-WFIRST15-0008, and Shota Rustaveli National Science Foundation of Georgia grants FR 19-498 and FR-19-8306. 

The authors are pleased to acknowledge that the work reported on in this paper was substantially performed using the Princeton Research Computing resources at Princeton University which is consortium of groups led by the Princeton Institute for Computational Science and Engineering (PICSciE) and Office of Information Technology's Research Computing.

We acknowledge the use of the NASA astrophysics data system \href{https://ui.adsabs.harvard.edu/}{https://ui.adsabs.harvard.edu/} and the arXiv open-access repository \href{https://arxiv.org/}{https://arxiv.org/}. Most numerical calculations presented in this manuscript were carried out using programming language \textsc{julia} \href{https://julialang.org/}{https://julialang.org/}. The software was hosted on the GitHub nplatform \href{https://github.com/}{https://github.com/}. The manuscript was typset using the overleaf cloud-based LaTeX editor \href{https://www.overleaf.com}{https://www.overleaf.com}.

\section*{Data Availability}
Computer codes used in producing the results presented in this manuscript\footnote{\href{https://github.com/ladosamushia/Bispectrum}{https://github.com/ladosamushia/Bispectrum}}\footnote{\href{https://github.com/ladosamushia/Bispectrum_SR}{https://github.com/ladosamushia/Bispectrum\_SR}} and instructions about \texttt{Quijote} access\footnote{\href{https://quijote-simulations.readthedocs.io/en/latest/}{https://quijote-simulations.readthedocs.io/en/latest/}} are available online.



\bibliographystyle{mnras}
\bibliography{higher_order_clustering}




\appendix

\section{Computing the bispectrum and its derivatives}
\label{app:technical_details}

We start by dividing each \texttt{Quijote} simulation box into a uniform grid of $N_\mathrm{grid}^3 = 512^3$ cells. We then go over all particles in the simulation and assign each cell a number depending on how far away the cell center is from the particle using the formula
\begin{equation}
    W = \begin{cases}
    (4 - 6s^2 + 3s^3)/6 \ &\textrm{for}\ 0 \leq s < 1\\
    (2 - s^3)/6 \ &\textrm{for}\ 1 \leq s < 2\\
    0\ &\textrm{for}\ 2 \leq s
    \end{cases},
\end{equation}
where $s$ is the distance between the particle and the cell center in units of the cell size. This assignment is based on a piecewise cubic spline prescription \citep{2004JCoPh.197..253C} and ensures that the gridding effects on the scales of interest (up to $k \sim 0.2h\ \mathrm{Mpc}^{-1}$) are negligible \citep{2016MNRAS.460.3624S}. Contributions from all particles are added up. This results in a $512^3$ array of numbers - $n_{ij\ell}$. We perform a discrete real-to-complex Fourier transform of $n_{ij\ell}$, to obtain a $512\times512\times257$ grid of complex numbers - $\widetilde{n}_{ij\ell}$, using the FFTW algorithm \citep{FFTW05}. The cells of this grid correspond to the wave-numbers at
\begin{align}
\nonumber
    k_x &= k_\mathrm{f}\left(i - \frac{N_\mathrm{grid}}{2}\right),\ \  i = 0,\ldots, N_\mathrm{grid},\\
    k_y &= k_\mathrm{f}\left(j - \frac{N_\mathrm{grid}}{2}\right),\ \ j = 0,\ldots, N_\mathrm{grid},\\
    \nonumber
    k_z &= k_\mathrm{f}\ell,\ \ \ \ \ \ \ \ \ \  \ell = 0,\ldots, N_\mathrm{grid}/2,\\
    k_\mathrm{f} &= \frac{2\pi}{L}
\end{align}
From this grid, we estimate the power spectrum $P(k)$ in bins of $\Delta k = 0.01 h\ \mathrm{Mpc}^{-1}$ starting with $k = 0$, and the bispectrum $B(k_1, k_2, k_3)$ in bins of $\Delta k_i = 0.01h\ \mathrm{Mpc}^{-1}$ starting with $k_i = 0$. 

For the power spectrum, we go over all cells in $\widetilde{n}_{ij\ell}$, compute the wave-number $k = \sqrt{k_x^2 + k_y^2 + k_z^2}$, and compute the absolute value of the entry, $|\widetilde{n}_{ijk}|^2$. For each $P(k)$ bin we simply average over contributions from all cells that fall into that bin.

For the bispectrum, we go over all pairs of cells in $\widetilde{n}_{ij\ell}$, find a third unique cell such that the ``triangular condition'' $\mathbf{k}_1 + \mathbf{k}_2 = \mathbf{k}_3$ holds for the three associated wave-numbers, and take a product $\widetilde{n}_1\widetilde{n}_2\widetilde{n}_3^\star$. The asterix in this equation denotes complex conjugation and subscripts refer to the triplet of integer indeces. For each $B(k_1, k_2, k_3)$ bin we take the average over contributions from all pairs of cells (and the third associated cell that is uniquely determined by the triangular condition) that fall into that bin. We do not look at the bispectrum bins unless at least some combination of the wave-numbers in that bin satisfies a triangular condition. To avoid counting equivalent triplets multiple times, we impose a condition $k_1 > k_2 > k_3$.
\begin{figure}
	\includegraphics[width=\columnwidth]{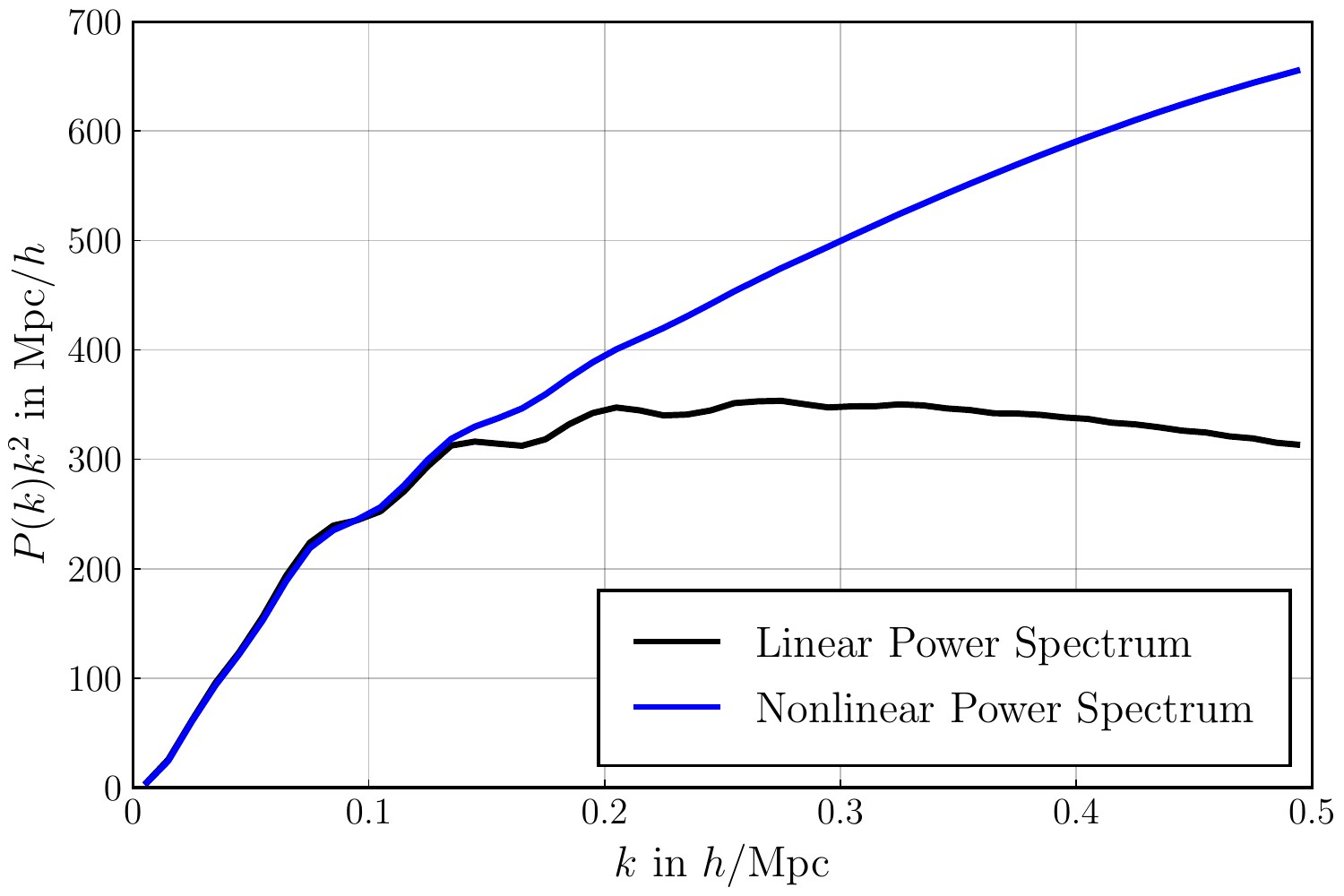}
    \caption{Nonlinear power spectrum at $z=0$ (blue) and a linear power spectrum (black) of \texttt{Quijote} matter particles. The linear power spectrum is measured from particles at $z=99$ and rescaled to have a matching amplitude on large scales.}
    \label{fig:Pk}
\end{figure}

Fig.~\ref{fig:Pk} shows the average nonlinear and linear power spectra of matter particles from \texttt{Quijote} simulations. We rescaled the linear power spectrum to have a matching amplitude at small wave-numbers. The two are reasonably close for large-scale modes (up to $k \sim 0.1h\ \mathrm{Mpc}^{-1}$ after which the additional power in the nonlinear power spectrum becomes visible. The partial erasure of the BAO signal is also visible by eye.
\begin{figure}
	\includegraphics[width=\columnwidth]{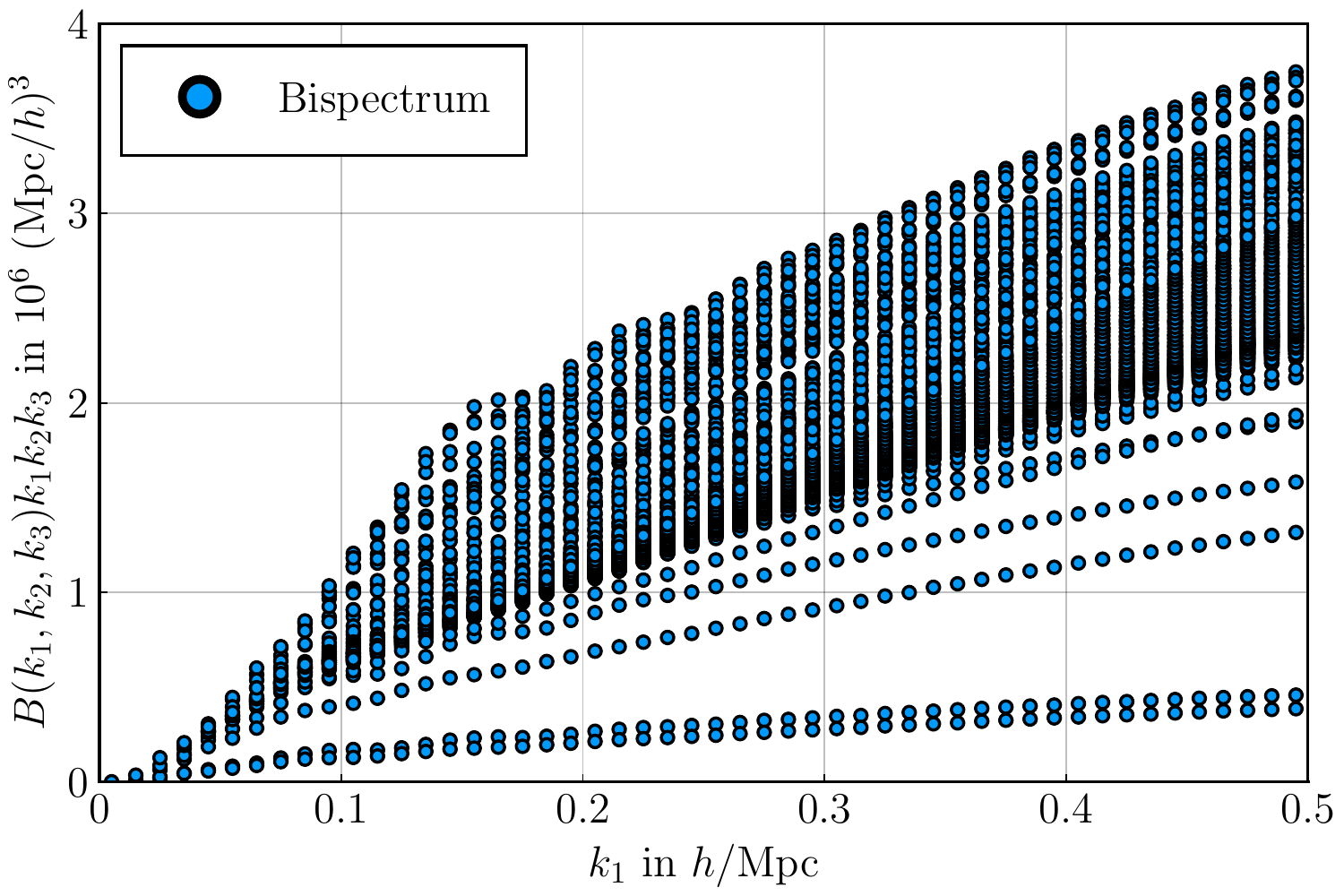}
    \caption{Bispectrum of \texttt{Quijote} matter particles at $z=0$ as a function of the largest wave-number $k_1$. Multiple points at the same abscissa show all bispectrum bins that share that value of $k_1$ but have different values of $k_2$ and $k_3$. We multiply the bispectrum by a factor of $k_1k_2k_3$. This factor is proportional to the number of fundamental triangles in the bin and separates different points for better visibility.}
    \label{fig:Bk}
\end{figure}
Fig.~\ref{fig:Bk} shows the average bispectrum of matter particles from \texttt{Quijote} simulations at $z=0$ as a function of the longest wave-number. Features in the shape of this three-dimensional function provide additional opportunities for the standard ruler test.

To estimate the covariance of these measurements, $\mathsf{C}$, by computing the sample covariance over 8,000 \texttt{Quijote} simulations,
\begin{align}
    \overline{X}_i &= \frac{1}{N_\mathrm{sim}}\displaystyle\sum_{\ell=1}^{N_\mathrm{sim}} X_{i\ell},\\
    C_{ij} &= \frac{1}{N_\mathrm{sim}}\displaystyle\sum_{\ell=1}^{N_\mathrm{sim}}  (X_{i\ell} - \overline{X}_i)(X_{j\ell} - \overline{X}_j).
\end{align}
$X_{i\ell}$ are the power spectrum and bispectrum measurements, where the first index goes over the bins and the second index goes over simulations. $N_\mathrm{sim}$ is the number of simulations.
\begin{figure}
	\includegraphics[width=\columnwidth]{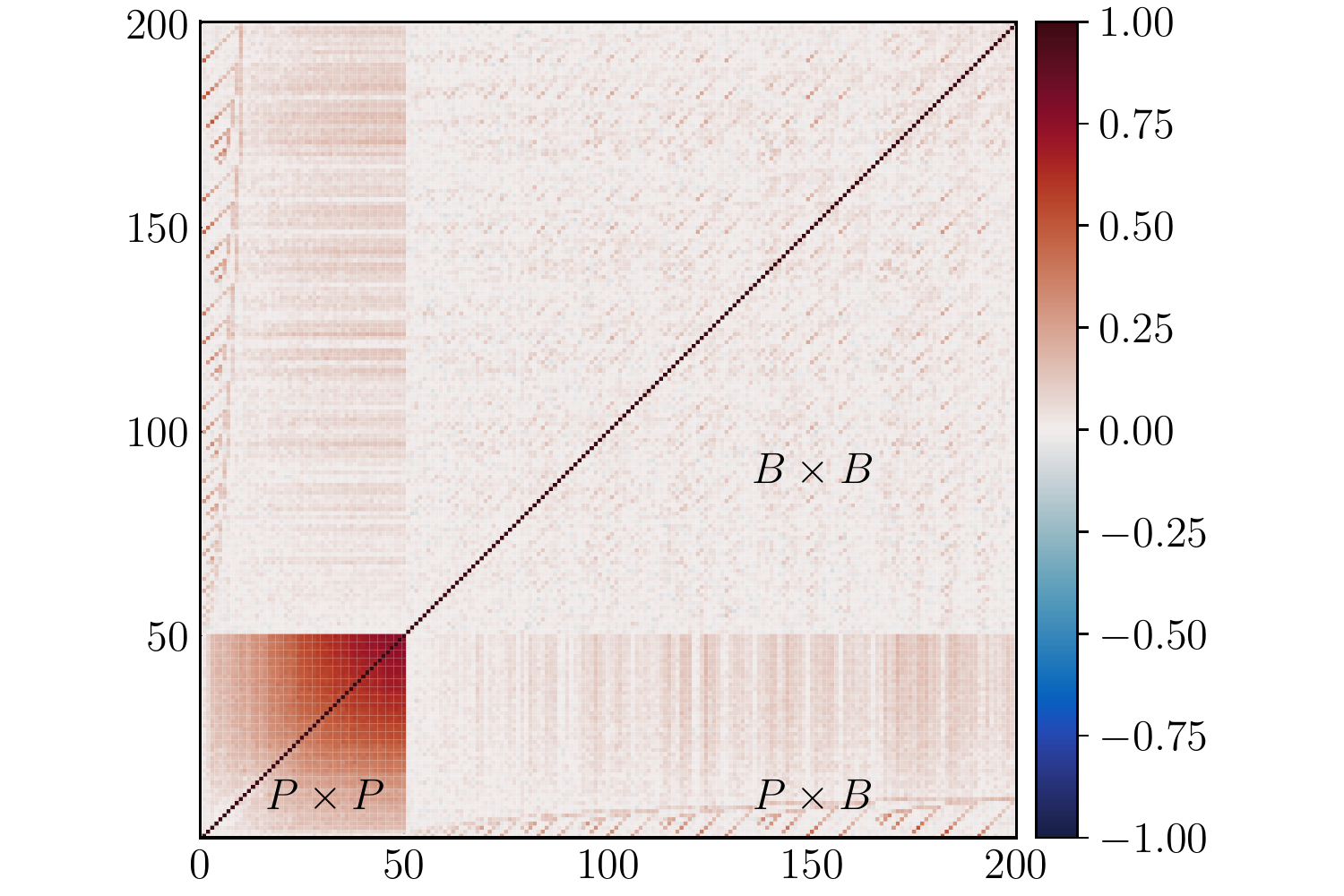}
    \caption{Correlation coefficient of the power spectrum (first 50 bins, goes up to $k_\mathrm{max} = 0.5h\ \mathrm{Mpc}^{-1}$) and the bispectrum (last 150 bins, goes up to $k_{1,\mathrm{max}} = 0.1h\ \mathrm{Mpc}^{-1}$) measurements. Colormap is adjusted in such a way that red, blue, and white cells correspond to the positive, negative, and zero correlations respectively.}
    \label{fig:rcov}
\end{figure}

Fig.~\ref{fig:rcov} shows the correlation matrix of these measurements defined as
\begin{equation}
    R_{ij} = \frac{C_{ij}}{\sqrt{C_{ii}C_{jj}}}.
\end{equation}
We ordered the measurements so that the first 50 elements are binned $P(k)$ up to $k_\mathrm{max} = 0.5h\ \mathrm{Mpc}^{-1}$. They are followed by binned $B(k_1,k_2,k_3)$ arranged in such a way that the bispectrum bin with smaller $k_1$ goes first (smaller $k_2$ when $k_1$ are equal, and smaller $k_3$ when both $k_1$ and $k_2$ are equal). The first 150 bins arranged in this way accommodate all bispectra up to $k_{1,\mathrm{max}} = 0.1h\ \mathrm{Mpc}^{-1}$. The correlation between power spectra at high wave-numbers is clearly visible in this plot. The bispectrum measurements are also correlated between themselves and the power spectra but to a lesser extent.

We apply a correction factor of 
\begin{equation}
    \mathcal{H} = \frac{N_\mathrm{sim}}{N_\mathrm{sim} - N_\mathrm{bins} - 1}
\end{equation}
to our inverse covariance matrices, where $N_\mathrm{bins}$ is the total number of power spectrum bin and bispectrum bin combinations. This factor corrects for the fact that the raw inverse of the sample covariance matrix tends to underestimate the errors \citep{2007A&A...464..399H,2013PhRvD..88f3537D}. This correction factor is the main reason we are unable to extend results presented on Figs.~\ref{fig:SNinfo} and \ref{fig:APinfo} to higher wave-numbers. The correction factor grows very rapidly with the $k_\mathrm{max}$, it is negligible for the most of the $k_\mathrm{max}$ but grows to 13 per cent at $k_\mathrm{max} = 0.2h\ \mathrm{Mpc}^{-1}$ for our choice of binning.

We compute the derivatives of power spectrum with respect to $\alpha$ by constructing a cubic spline interpolation over power spectrum bins and then taking a backward numeric derivative,
\begin{equation}
    \frac{\mathrm{d}P}{\mathrm{d}\alpha} = \frac{P(k) - P(k(1 - \epsilon))}{\epsilon},
\end{equation}
with $\epsilon=1\times10^{-6}$. To use a forward numeric derivative,
\begin{equation}
    \frac{\mathrm{d}P}{\mathrm{d}\alpha} = \frac{P(k(1+\epsilon)) - P(k))}{\epsilon}.
\end{equation}
We use forward numerical differentiation for the lowest $k$ value for which the backward differentiation would require an extrapolation. The bispectrum derivatives are computed similarly. We first construct a three-dimensional cubic spline and then compute a numeric derivative,
\begin{equation}
     \frac{\mathrm{d}B}{\mathrm{d}\alpha} = \frac{B(k_1,k_2,k_3) - B(k_1(1-\epsilon),k_2(1-\epsilon),k_3(1 - \epsilon))}{\epsilon},
\end{equation}
except for some bispectra bin combinations with small values of $k$ where backwards numerical differentiation would require extrapolating our spline fit.

\section{Standard Rulers}
\label{app:standardruler}

We need to assume a fiducial cosmological model to convert galaxy redshifts to comoving distances $d_\mathrm{c}$,
\begin{equation}
    d_{\rm{c}} = d_\mathrm{c}(z, \boldsymbol{\theta}).
\end{equation}
The measured distances between galaxies will then be different from the real distances by
\begin{align}
    \left|\mathbf{r}_1 - \mathbf{r}_2\right| &= \left|\mathbf{r}_1 - \mathbf{r}_2\right|_\mathrm{measured}\ \frac{d_\mathrm{c}(z, \boldsymbol{\theta}_\mathrm{fid})}{d_\mathrm{c}(z, \boldsymbol{\theta})} \\
    \nonumber
    &\equiv \left|\mathbf{r}_1 - \mathbf{r}_2\right|_\mathrm{measured}\ \alpha(z, \boldsymbol{\theta}_\mathrm{fid}, \boldsymbol{\mathrm{\theta}}),
\end{align}
where $d_{\mathrm c}$ is the physical distance to those galaxies. In conventional cosmological models this distance is given by
\begin{equation}
    d_\mathrm{c}(z) = c\displaystyle\int_0^z\frac{dz'}{H(z')},
\end{equation}
where $c$ is the speed of light and $H(z)$ is the Hubble expansion parameter as a function of redshift.
When fitting models of the correlation function $\xi$ to the measurements we have to account for this scale-independent dilation,
\begin{equation}
    \xi(r,\boldsymbol{\theta}) = \xi(r\alpha(\boldsymbol{\theta}), \boldsymbol{\theta}).
\end{equation}
The dilation effect is similar for the higher-order NPCFs, but we will keep our discussion to the 2PCF for simplicity. The power spectrum, being the Fourier transforms of the 2PCF, scales by the inverse factor,
\begin{equation}
    P(k,\boldsymbol\theta) = P(k/\alpha(\boldsymbol{\theta}),\boldsymbol{\theta}).
\end{equation}
Rather then being a nuisance, the presence of this scaling provides additional opportunities for constraining cosmological parameters. Small changes in dark energy parameters, for example, affect the intrinsic shape of the power spectrum (the second argument) weakly, but affect the distance-redshift relationship and consequently the dilation of the shape strongly. Recent works suggested that it may be more natural to make NPCF measurements directly in the observable space of redshifts and angles \citep{1999MNRAS.305..527T,2011PhRvD..84f3505B,2014PhRvD..90f3515N,2018JCAP...04..029Y,2020PhRvL.124c1101J}. In this approach, the dilation is absent from the measurements (which are made in redshift and angular coordinates) and the distance-redshift relationship becomes part of the model. We present our arguments in a more familiar setting, because then, it makes the usage of the polyspectra as standard rulers more explicit. Our results would not change if we instead formulated our arguments in terms of observational coordinates (redshifts and angles).

Small uncertainties in the measured power spectrum propagate to inferred cosmological parameters as
\begin{equation}
    \delta P(k) \sim -\frac{\partial P(k)}{\partial k}\frac{\partial\alpha}{\partial\boldsymbol{\theta}}\frac{k}{\alpha^2}\delta\boldsymbol{\theta} + \frac{\partial P(k)}{\partial\boldsymbol{\theta}}\delta\boldsymbol{\theta}.
\end{equation}
Fitting only the BAO feature in the power spectrum makes the separation between the standard ruler and intrinsic shape constraints clearer. The power spectrum can be divided into BAO and smooth components,
\begin{equation}
    P(k, \boldsymbol{\theta}) = P_\mathrm{BAO}(k, \boldsymbol{\theta}, \boldsymbol{\sigma})P_\mathrm{smooth}(k, \boldsymbol{\theta}, \boldsymbol{\nu}),
\end{equation}
where by $\boldsymbol{\sigma}$ and $\boldsymbol{\nu}$ we denoted nuisance parameters needed in real analyses to compensate for observational effects and inaccuracies of theoretical modeling
\citep[see e.g. ][for details of how this split is performed in practice]{Eisenstein_1998_NW}. $\boldsymbol{\sigma}$ will include parameters that describe the nonlinear damping of the BAO feature, which is difficult to link to the cosmological parameters $\boldsymbol{\theta}$ directly, and $\boldsymbol{\nu}$ will include smooth polynomials in $k$ that describe the effect of galaxy bias and nonlinear evolution (also difficult to compute from $\boldsymbol{\theta}$ based on first principles). Both components will dilate by $\alpha$ but since the second component is ``smooth'' the effects of dilation will be degenerate with the  nuisance parameters. The BAO part of the power spectrum has a decaying oscillatory feature the dilation of which cannot be mimicked by nuisance parameters. For most conventional cosmological models, the $P_\mathrm{BAO}$ between two models can be matched by rescaling the power spectrum by a factor of $r_\mathrm{d}(\boldsymbol{\theta})$ and adjusting nuisance parameters $\boldsymbol{\sigma}$ as in
\begin{equation}
    P_\mathrm{1,BAO}\left[r_{1,\mathrm{d}}(\boldsymbol{\theta}_1)k,\boldsymbol{\sigma}_1\right] = P_\mathrm{2,BAO}\left[r_{2,\mathrm{d}}(\boldsymbol{\theta}_2)k,\boldsymbol{\sigma}_2\right],
\end{equation}
where subscripts 1 and 2 denote two different cosmologies. $r_\mathrm{d}$ can be accurately computed in each cosmology. For these BAO only fits then measuring a power spectrum in a fiducial cosmology and comparing it to a fiducial shape through dilation allows us to measure the combination $r_\mathrm{d}/\alpha$ through
\begin{equation}
    P(kr_\mathrm{d,fid}) = P(\alpha kr_\mathrm{d}).
\end{equation}
The measurement of $\alpha$ can then be interpreted as a measurement of
\begin{equation}
    \alpha(\boldsymbol{\theta}) = \frac{d_\mathrm{c}(z,\boldsymbol{\theta})r_\mathrm{d,fid}}{r_\mathrm{d}(\boldsymbol{\theta})d_\mathrm{c,fid}}.
\end{equation}

Sensitivity of measured power-spectrum to $\alpha$ is determined by its log-derivative with respect to the wave-number,
\begin{equation}
    \delta\alpha \sim \left[\frac{\partial P(k)}{\partial \ln k}\right]^{-1}\delta P(k).
\end{equation}
All other things equal a feature at high wave-numbers would provide a better constraint of $\alpha$. The power spectrum (or higher order spectra) can be used as standard rulers as long as they are not flat in $k$ in which case they have no sensitivity to dilation. Parts of the power spectrum that scale as pure power law also do not contribute to the $\alpha$ constraints since they dilation in this case is fully degenerate with the linear bias parameter. Linear bias cannot be determined from cosmological parameters and would have to be measured from the power spectrum along with $\alpha$. To see why this occurs, consider measuring the power spectrum at $N$ different wave-number bands centered around $k_i$, where $i=1\ldots N$, and in that in this wave-number range the power spectrum can be modeled as $P_i = b^2k_i^n$. The derivatives of the power spectrum with respect to $\alpha$ and unknown bias will then be,
\begin{align}
    \frac{\partial P_i}{\partial \alpha} = nb^2k_i^n,\ \ \
    \frac{\partial P_i}{\partial b} = 2bk_i^n.
\end{align}
These derivative vectors are linearly dependent; therefore for the pure power-law power spectrum, the constraints on $\alpha$ will be fully degenerate with the constraints on the amplitude.


\bsp	
\label{lastpage}
\end{document}